%% file: main.tex
\documentclass[conference]{IEEEtran}
\IEEEoverridecommandlockouts

\usepackage[inline,shortlabels]{enumitem}
\usepackage{comment}
\usepackage{algpseudocode}
\usepackage{algorithm}
\usepackage{graphicx}
\usepackage{pgfplots,pgfplotstable}
\usepackage{xparse}
\usepackage{subcaption}
\usetikzlibrary{positioning,calc}
\usepackage{amsmath}
\usepackage{amsfonts} 

\pgfplotsset{compat=1.18}
\usepackage{listings}
\usepackage{booktabs}
\usepackage[many]{tcolorbox} 
\usepackage{stfloats}
\usepackage{microtype}
\usepackage{fancyhdr}
\usepackage{url}

\RequirePackage[
  style=acmnumeric,
  maxnames=2, 
  minnames=1, 
  ]{biblatex}

\usepackage[hidelinks]{hyperref}
\usepackage[capitalise]{cleveref}

\fancypagestyle{fancycopyright}{%
    \fancyhf{}%
    \fancyhead[C]{Author copy of paper published at \emph{17th IEEE/ACM International Conference on Utility and Cloud Computing (UCC 2024)}\\
    \copyright2024 IEEE 
    }%
    \fancyfoot[C]{\thepage}%
}

\include{plots/pgplots-definitions}
\begin{document}

\title{Truffle: Efficient Data Passing for Data-Intensive Serverless Workflows in the Edge-Cloud Continuum}

\author{
        \IEEEauthorblockN{Cynthia Marcelino}
        \IEEEauthorblockA{\textit{Distributed Systems Group, TU Wien} \\
            c.marcelino@dsg.tuwien.ac.at
        }
        \and
        \IEEEauthorblockN{Stefan Nastic}
        \IEEEauthorblockA{\textit{Distributed Systems Group, TU Wien} \\
            snastic@dsg.tuwien.ac.at}
}

\maketitle


\thispagestyle{fancycopyright}
\pagestyle{fancy}

\begin{abstract}
Serverless computing promises a scalable, reliable, and cost-effective solution for running data-intensive applications and workflows in the heterogeneous and limited-resource environment of the Edge-Cloud Continuum.
However, building and running data-intensive serverless workflows also brings new challenges that can significantly degrade the application performance. 
Cold start remains one of the main challenges that impact the total function execution time. Further, since the serverless functions are not directly addressable, Serverless workflows need to rely on external (storage) services to pass the input data to the downstream functions. Empirical evidence from our experiments shows that the cold start and the function data passing take up the most time in the function execution lifecycle. 

In this paper, we introduce Truffle -- a novel model and architecture that 
enables efficient inter-function data passing in the Edge-Cloud Continuum by introducing mechanisms that separate computation and I/O, allowing serverless functions to leverage the cold starts to their advantage. 
Truffle introduces Smart Data Prefetch (SDP) mechanism that abstracts the retrieval of input data for the serverless functions by triggering the data retrieval from the external storage during the function's startup.
Truffle's Cold Start Pass (CSP) mechanism optimizes inter-function data passing and data exchange within serverless workflows in the Edge-Cloud Continuum by hooking into the functions' scheduling lifecycle to trigger early data passing during the function's cold start.
Experimental results show that by leveraging the data prefetching and cold-start data passing, Truffle reduces the IO latency impact on the total function execution time by up to 77\%, improving the function execution time by up to 46\% compared to the state-of-the-art data passing approaches.

\end{abstract}


\section{Introduction}

Serverless computing is a paradigm where short-lived functions are executed in response to an external triggering event. Once the trigger event is received, the Serverless platform provisions the necessary environment infrastructure for the function to execute. The function provisioning at the startup of the function lifecycle is also known as cold start~\cite{whatServerlessIsAndWhatShouldBecome,WhereWeAreLiesAhead}. 
Despite efforts in minimizing it~\cite{SEUSS,Xanadu}, cold starts represent a significant proportion of the function's total execution time, reaching up to 80\%~\cite{faascache, FaaSLight}. In Serverless computing, where functions are short-lived, over 50\% have execution times under 100ms. Thus, leveraging the longest tasks of the serverless function lifecycle, such as cold starts, becomes crucial for improving the overall function execution time~\cite{atoll, state-of-serverless}.

The stateless design of Serverless Computing pushes functions to rely on external systems for data exchange, with storage services representing over 60\%, and messaging services comprising 38\%. Furthermore, 30\% of the functions process data over 10 MB, whereas 26\% are over 100MB~\cite{state-of-serverless,serverless-why-when-how}. Nevertheless, using external services to pass data adds significant network and latency overhead and consequently increases the function end-to-end execution~\cite{sonic,sand,cloudburstSF}. Moreover, external services introduce additional development effort as applications must explicitly fetch their ephemeral input data~\cite{pocket, Lambdata}. 
Therefore, constant efforts are necessary to improve the data passing between serverless functions. 

The most common approaches for data passing in serverless functions include: 
\begin{enumerate*} [label=(\alph*)]
    \item \textit{Remote storage services}, such as object storage~\cite{s3}, Key-Value Store  (KVS)~\cite{pocket, cloudburstSF} and  cache~\cite{InfiniCache,duo} are the most common mechanism for data passing. Nevertheless, it increases network overhead and latency up to 95\%~\cite{faastlane}; 
    \item \textit{Local storage:} approaches such as disk storage~\cite{sonic,goldfish2024}, shared memory~\cite{faasm,nightcore} and local cache~\cite{ofc,cloudburstSF} shift the data storage to local workers. It enables serverless platforms to exploit function and data locality to improve latency and throughput for co-located functions. Thus, decreasing remote communication. 
\end{enumerate*}
Unfortunately, these approaches do not consider the function cold start in the data passing mechanisms.

Furthermore, cold starts remain a challenge in Serverless computing. The Serverless platform provisions the full infrastructure for function execution. During the cold start phase, the infrastructure is not fully operational, leading to idle waiting time for both the function and the host. Consequently, the cold start influences the overall duration of end-to-end function execution~\cite{cold_start_inserverless,Prebaking}. 

Common approaches that address cold start include:
\begin{enumerate*} [label=(\alph*)]
\item \textit{Caching} solutions propose to store full or partial function sandbox snapshots in memory for a certain time. 
Serverless platforms create a snapshot of the function sandbox to reuse it in new requests. Thus, Serverless platforms only provision the language runtime instead of the language runtime and infrastructure~\cite{faascache,IceBreaker}. 
Although caching decreases the cold start latency significantly, it increases resource usage~\cite{Retention-Aware};
\item \textit{Sandbox sharing} approach relies on multiple functions running in one sandbox. As functions run in a shared sandbox, platforms profit from a single cold start for multiple functions, reducing cold start latency and optimizing function execution time~\cite{sand,SOCK,Cwasi2023}. 
\item \textit{Sandbox techniques} such as tiny VM and Web Assembly VM propose minimal sandboxes with only the necessary components and features required by the function. Additionally, it offers an additional layer of security. Thus, sandbox techniques decrease the resource provisioning overhead while increasing security~\cite{faasm,Firecracker}.
\end{enumerate*}

The presented approaches optimize serverless functions cold starts by addressing the function preparation and setup during the function initialization. Nevertheless, cold start solutions typically overlook the crucial phases of the function lifecycle, such as data transfer which along with the cold start have a significant impact on the Serverless function execution time.

In this paper we introduce Truffle, a modular architecture that enables input data fetching and passing. Truffle's communication technique leverages cold starts to transfer input data to another function efficiently. 
We can summarize our contributions as follows:

\begin{itemize}
    \item \textbf{Truffle}, a novel model and architecture that separates computation and I/O to enable functions to execute tasks in parallel to the functions' cold starts for optimized end-to-end function execution. 
    
    \item \textbf{The Smart Data Prefetch (SDP) mechanism}, a function input data fetching mechanism that abstracts the retrieval of input data for individual functions. Truffle seamlessly identifies the input data storage type and triggers the data retrieval at cold start, i.e., before function execution. 
    
    \item \textbf{The Cold Start Pass (CSP) mechanism}, an inter-function data passing mechanism that optimizes data exchange between serverless functions. It identifies the function target host immediately after the scheduling phase preceding function execution. Truffle utilizes the cold start to transmit the data, optimizing inter-function communication. Consequently, when the target function executes, the input data is already stored in a buffer close to the function.
\end{itemize}

Truffle reduces the IO latency impact on the function execution time by up to 77\%, achieving up to 46\% reduction of the overall function execution time by leveraging the cold start to prefetch and pass the data to another function. Moreover, applications with longer cold starts of up to 10s profit nearly 30\% more from Truffle compared to applications with short cold starts of up to 2s.

\section{Motivation \& Illustrative Scenario} \label{motivation}

\subsection{Motivation}

Serverless Computing is typically characterized by \emph{stateless} and \emph{non-addressable} functions deployed on entirely managed platforms. In the Edge-Cloud Continuum, Serverless computing enables efficient function execution on resource-limited edge devices while leveraging the Cloud for computing intense tasks, thus optimizing performance and reducing latency across distributed environment of the Edge-Cloud Continuum. 
By design, serverless functions lack direct addressability. Instead, these functions are accessible to clients through platform ingresses such as Load Balancers and API Gateways~\cite{scf, CloudProgrammingSimplified}. Furthermore, Serverless platforms limit their function input data size (AWS $\lambda$: 6MB, OpenFaas: 1MB, GCP: 10MB), pushing functions to leverage external services such as object storage, KVS and message queues to pass larger amount of data which not only introduces complexity in the data handling process but also adds significant latency due to the reliance on external services~\cite{sonic,onestep}. 

\begin{figure}[t]
\begin{tikzpicture}
\pgfplotstableread{ 
Label    Sched  Prep     Transfer   Exec
D-128MB   20     2375     1291       15  
KVS-128MB 16     2033     1584       12  
S3-128MB  34     1660     2481       15  
}\datatable

\begin{axis}[scale=.9,
    xbar stacked,   
    xmin=0,         
    ytick=data,     
    yticklabels from table={\datatable}{Label},
    legend style={area legend, at={(0.36,1.7)}, anchor=north, draw=none,legend columns=-1},
    ymajorgrids=true,
    xmajorgrids=true,
    xlabel=Latency in ms,
    height=3cm,
    width=9cm
]
\addplot [fill=yellow] table [x=Sched, y expr=\coordindex] {\datatable};    
\addplot [fill=blue!50]table [x=Prep, y expr=\coordindex] {\datatable};
\addplot [fill=cyan!70!yellow] table [x=Transfer, y expr=\coordindex] {\datatable};
\addplot [fill=orange] table [x=Exec, y expr=\coordindex] {\datatable};
\legend{Scheduling,Cold Start, Data Transfer,Fn Execution}
\end{axis}
\end{tikzpicture}
\caption{Knative Serverless function lifecycle steps duration with data passing approaches Direct (D), Key-Value Store (KVS), and Object Storage (S3) for 128MB}
\label{fig:coldstart}
\end{figure}
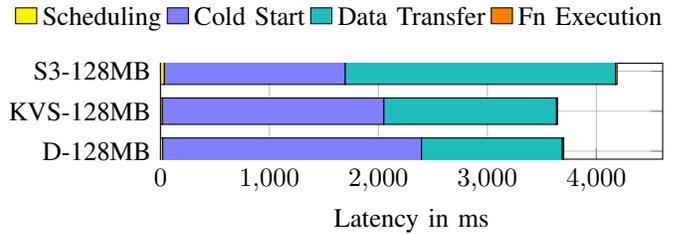

\cref{fig:coldstart} shows the function latency of each step in the Serverless function lifecycle with the most common storage types, such as AWS S3, KVS, and Direct. Experimental evidence indicates that cold start and data transfer constitute a significant portion of the function execution time, and combined, they contribute to up to 99\% of the function latency. 

\begin{figure}[!ht]
\centering
\includegraphics[width=.8\linewidth]{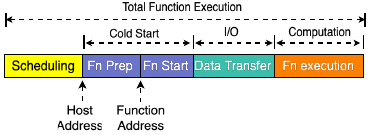}
\caption{Serverless Function Lifecycle with 1MB Input Data Transfer}
\label{fig:fh_lifecycle}
\end{figure}

\cref{fig:fh_lifecycle} shows the four phases of function lifecycle: scheduling, cold Start, I/O and computation. Moreover, \cref{fig:fh_lifecycle} shows that the function host is known immediately after the \textit{Scheduling} phase. Nevertheless, the \textit{Data Transfer} (I/O) only initiates after \textit{Fn Start} when the Serverless function executes. In summary, the function and host remain idle until the serverless function provision has been completed. 
As shown in \cref{fig:fh_lifecycle}, the function \textit{Cold Start} and \textit{Data Transfer} contribute significantly to the function's end-to-end execution time. Moreover, the function is only available to process the request after \textit{Fn Start} when the infrastructure has finished and the language runtime is fully running. However, the host is known after the \textit{Scheduling}, implying that I/O tasks such as data transfer may initiate immediately after scheduling in parallel to the cold start. Despite this parallelization potential, the delays caused by cold starts and data transfers remain substantial, presenting challenges for optimizing performance in latency-sensitive applications such as data-intensive Serverless workflows in the Edge-Cloud Continuum.

\subsection{Illustrative Scenario}

To better motivate our challenges, we present an illustrative scenario for real-time video analytics that focuses on detecting fire emergencies in smart cities. To achieve this, cameras and sensors are strategically positioned throughout the city to detect fire patterns. A Serverless workflow is employed to identify and respond to fire emergencies.

\begin{figure}[!b]
\centering
\includegraphics[width=.7\linewidth]{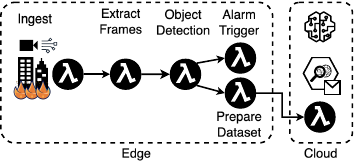}
\caption{Simplified Serverless Workflow for Fire Detection for Smart Cities}
\label{fig:env_monitoring}
\end{figure}

Our workflow utilizes five Serverless functions, partially executed on the Edge and partially executed on the Cloud. To reduce communication latency in our workflow, some tasks are executed at the Edge, close to the data source. Edge tasks are responsible for processing large real-time video streams, extracting image frames, simple object detection, and triggering immediate local alerts for emergencies. On the other hand, tasks that require more powerful computing resources, such as more complex object detection and model training, are carried out in the Cloud. Our motivating scenario is inspired by a Serverless Workflow for real-time environmental monitoring~\cite{Escobar2024UnleashingTP,tema,yolo}.

In \cref{fig:env_monitoring}, in \textit{Ingest} stage, real-time videos captured by cameras are transmitted to edge nodes via a streaming framework, where serverless functions responsible for \textit{Extract Frames} are activated to process the video data in small chunks, effectively reducing latency. Each \textit{Extract Frames} function processes a video segment, ensuring swift data handling. Upon completing their tasks, these functions directly pass the processed frames to the \textit{Object Detection} functions, who analyze them to identify specific fire patterns such as smoke and flames. Following the detection process, the \textit{Object Detection} functions communicate directly with the \textit{Alarm Trigger}  functions, who evaluate the data to decide whether to trigger local emergency responses. Concurrently, \textit{Object Detection} functions send data to \textit{Prepare Dataset} functions for data preprocessing. Finally, the processed data is transmitted to the cloud, where more resource-intensive tasks are performed, such as training machine learning models to enhance fire detection. 

Truffle enhances Serverless workflows by separating computation and I/O, which allows Truffle to prefetch and transfer the input data during cold starts. Decreasing function latency enables quick data processing, which is crucial for real-time applications like fire detection and emergency response in the Edge-Cloud Continuum. Truffle optimizes data transfer between edge devices and cloud servers, ensuring data is available for computation as soon as the functions are initialized. By separating I/O from computation, Truffle can incorporate the input data transfer into the function startup phase, facilitating faster and more efficient data handling, essential for maintaining data-intensive Serverless workflows in the Edge-Cloud Continuum.

\section{Truffle Model \& Architecture Overview} \label{sec2}

\subsection{Truffle Model} \label{subsec:truffle_model}

The execution time $\tau$ of a Serverless function can be modeled by the sum of its individual tasks. Given a Set of functions $\lambda$ with function index $i$, where $i \in \mathbb{N}$. Thus, the cold start $\beta_i$ can be defined as: 

\begin{equation}
    \beta_i = \upsilon_i + \eta_i 
\end{equation}
Where $\upsilon$ represents the function infrastructure setup and $\eta$ the function startup. Cold starts introduce additional steps such as scheduling, infrastructure setup, and language runtime startup, resulting in increased latency. 
Consequently, cold starts directly influence the overall latency of end-to-end functions. Additionally, in the state-of-the-art platforms, every step of the Serverless functions cold starts are executed in sequence.
Truffle proposes to execute the cold and the data transfer in parallel to optimize the function latency. Thus, the improvement $\varphi_i$ is defined as follows:

\begin{equation}
\varphi_i = \max(\beta_i, \delta_i)
\end{equation}

Where $\beta_i$ is the function cold start and $\delta_i$ is the data transfer time. Since Truffle executes the cold start and data transfer simultaneously, $\varphi$ is defined by the longest task between these two tasks, i.e., cold start and data transfer.

Therefore, the end-to-end execution time $\tau$ of a Set of functions $\lambda$ can be described as follows:

\begin{equation}
    \tau (\lambda_i) = \sum_{i=1}\alpha_i + \max((\upsilon_i + \eta_i), \delta_i) + \gamma_i 
\end{equation}

Here, the total execution time of a function $\lambda_i$ is equal to the scheduling $\alpha_i$ plus the maximum value between the sum of function infrastructure setup $\upsilon_i$ + function startup $\eta_i$ and data transfer $\delta_i$. Thus, $\gamma_i$ represents the function execution time ~\cite{sonic,serverlessTheory,COCOA,HyperDrive2024}. 
Therefore, we can define truffle improvement $\Delta_i$ in the end-to-end execution time as follows:

\begin{equation}
    \Delta_i = (\beta_i+\delta_i) - \max(\beta_i, \delta_i)
\end{equation}

Finally, the Truffle optimization model aims to minimize the end-to-end execution of a Serverless workflow composed of a Set of serverless functions $\lambda$. To achieve that, Truffle considers three main phases for each function: the scheduling time \(\alpha_i\), the longer time between the cold start and data transfer \(\max(\beta_i, \delta_i)\), and the function execution time (\(\gamma_i\)) as follows: 

\begin{equation}
\min \sum_{i=1}^{n} \left( \alpha_i + \max(\beta_i, \delta_i) + \gamma_i \right)
\end{equation}

Hence, Truffle improvements are directly connected to the cold start latency and data input size. Longer cold starts associated with longer data transfer have more profit than shorter data transfer and cold starts.

\subsection{Truffle Architecture Overview}

\begin{figure}[!b]
\centering
\includegraphics[width=.9\linewidth]{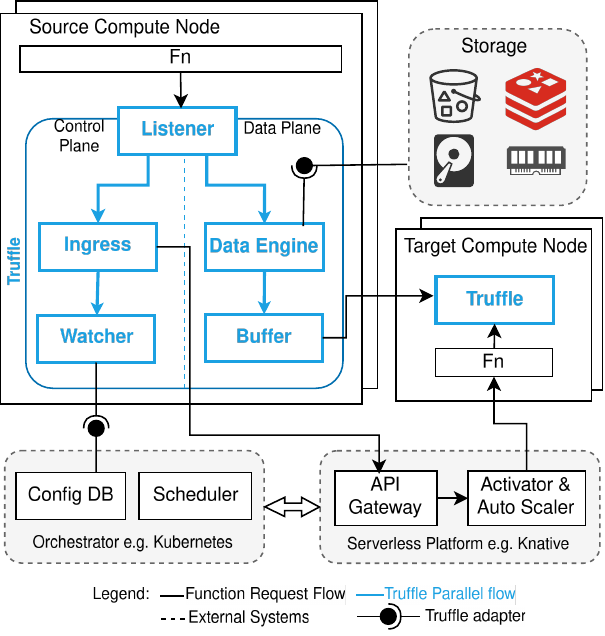}
\caption{Truffle Architecture Overview}
\label{fig:truffle_overview}
\end{figure}

In the state-of-the-art platforms, I/O tasks such as data transfer initiates after fully provisioned function. Furthermore, as Serverless functions are sensitive to network latency, the function execution time grows proportionally to the network data transfer~\cite{sonic,sand,fnAsFunction}.  

Truffle is designed to optimize the data input handling for serverless functions. It introduces a novel architecture to identify, retrieve, and transfer the data input based on function properties such as storage type, event type, serverless platform, and orchestration tools.
Truffle proposes to separate the computation and I/O so that the network data input transfer can start simultaneously with the cold start process right after scheduling. This means that the functions that are currently running, also known as hot functions, may not benefit from Truffle. In such cases, Truffle works as a proxy and only pass the data through it.

Despite extensive efforts to mitigate cold starts in serverless architectures, the issue persists and can significantly impact the total function execution. By enabling functions to transfer data during cold starts, typically an idle time for the serverless function, Truffle reduces the function latency while alleviating serverless functions' burden of fetching input data.
Truffle is located on each computing node to intercept local requests to improve inter-function communication latency. Additionally, it uses orchestration tools like Kubernetes for better scalability and throughput.

\cref{fig:truffle_overview} shows Truffle's interaction with the function and the serverless and orchestration platform. Truffle has a modular architecture that allows each component to be extended and eventually replaced. Truffle components enable two main mechanisms: Smart Data Prefetch and Cold Start Data Pass. 
\textit{Smart Data Prefetch}, detailed in ~\cref{sec3:data_fetching}, is triggered at every function start and retrieves the function input data from the respective storage such as direct, object storage, and KVS in parallel to the cold start. Whereas \textit{Cold Start Data Pass}, described in ~\cref{sec3:data_passing}, facilitates and optimizes the data passing between Serverless functions.

\subsubsection{Truffle Components}
Truffle is composed of five components: Listener, Data Engine, Ingress, Watcher, and Buffer.

\paragraph{Listener} It connects to the Serverless functions invoking events. The goal of the listener is to intercept the function request trigger and notify Truffle of an incoming request. It supports multiple serverless functions invoking events such as HTTP requests, streams, queues, and jobs. 

\paragraph{Ingress} It concurrently triggers the Serverless platform and Truffle's Data Engine to initiate the function input data fetching. More specifically, the ingress accepts requests from the source client functions. Then, it sends an event to the Serverless platform with a key reference to the function input data stored in Truffle's local buffer. Subsequently, the serverless platform is responsible for forwarding the request and implementing the function's scaling procedures if necessary.   

\paragraph{Data Engine} It identifies the storage type of incoming function data, retrieves it, and stores it in the buffer. This component enables functions to seamlessly receive input data, regardless of their input data storage type, such as KVS and object storage. Truffle's architecture design with adapters facilitates the extension of the data engine, supporting various storage types and multiple cloud providers. Consequently, it increases flexibility and facilitates serverless function development.

\paragraph{Watcher} It connects to the underlying orchestration tool, such as Kubernetes via the control API and listens for live events, such as scheduling events and function host assignments. Its primary task is identifying the assigned host for a target function, allowing Truffle to pass data between functions. The watcher recognizes the data input source and notifies Truffle ingress once the target function has been assigned to a worker or if the function already has an assigned worker, indicating that the function is already running. In cases of hot function, Truffle does not interfere and forwards the request without any modification. Moreover, the watcher employs an adapter to communicate with multiple orchestration tools, such as Kubernetes and Nomad.

\paragraph{Buffer} It stores the data until the target function is fully provisioned. The buffer can be placed on each node locally or remotely via external services. A local buffer keeps the data closer to the function and enables high-speed access to the data via local storage, such as in-memory storage. The buffer can also leverage remote services, such as databases and KVS, to provide flexibility and scalability. In cases where local computing resources are limited, the remote buffer enables functions to leverage the cold start for data passing with on-demand auto-scaling without local resource overhead.

\section{Truffle Runtime Mechanisms} \label{sec3}


\subsection{The Smart Data Prefetch}\label{sec3:data_fetching}

\begin{figure}[!b]
\centering
\includegraphics[width=.8\linewidth]{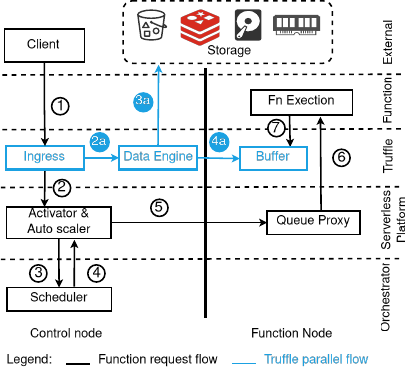}
\caption{Truffle Smart Data Prefetch}
\label{fig:truffle_datafetching}
\end{figure}

Truffle Smart Data Prefetch (SDP) abstracts and optimizes the data fetching from the serverless function. 
\cref{fig:truffle_datafetching} shows how Truffle fetches the function input data. \textcircled{\small{1}} First, Truffle ingress receives the client request with all the required information to retrieve the input data as arguments to the function handler, such as the bucket name, object ID, and credentials. 
\textcircled{\small{2}} Truffle ingress sends two simultaneous requests. One request to the platform Activator \& Auto Scaler and one request to the Data Engine.
\textcircled{\small{2a}} At this point, Truffle initiates an asynchronous request path parallel to the original request flow, e.g., requests via serverless platforms. Truffle triggers the Data Engine component to identify and start the data fetching. 
\textcircled{\small{3}} The Auto Scaler scales up if necessary, and the scheduler gets then notified about new scheduling.
\textcircled{\small{3a}} Truffle initiates the input data fetching from the respective storage, i.e., direct, object storage, or KVS. For specific cases where special access and roles are requested, such as AWS S3 credentials and keys, the function must specify and give Truffle the right access via specific function properties.
\textcircled{\small{4}} After scheduling, The Activator \& Auto Scaler get notified the scheduling has finished and the request may resume.
\textcircled{\small{4a}} Once the data fetching is finished in the Data Engine, Truffle stores the input data size in the Buffer. 
\textcircled{\small{5}} The Activator resumes the request and forwards it to the Serverless Platform Queue Proxy.
\textcircled{\small{6}} The Queue Proxy forwards the request to the Serverless function handler during Fn Execution.
\textcircled{\small{7}} Once the function executes in Fn Execution, it retrieves its input data from the Truffle Buffer.

\begin{algorithm}[!tb]
\caption{Truffle Smart Data Prefetch}
\label{alg:datafecher}
\begin{algorithmic}[1]
\Require $R$: incoming request
\Require $B_N$: buffer name
\Require $S_T$: storage type
\Require $C_R$: content reference
\State $B$ $\leftarrow$ buffer
\State $SC$ $\leftarrow$ storage
\ForAll { $B_T \in BC$ }
    \If{$ B_T \ne \emptyset \land B_T = B_N$ }
        \State $B \leftarrow B_T$
    \EndIf
\EndFor
\ForAll { $S_E \in SC$ }
    \If{$ S_E \ne \emptyset \land S_E = S_T$ }
        \State $SC$ $\leftarrow$ getclient($S_T$)
    \EndIf
\EndFor
\State $C \leftarrow SC.get(C_R))$
\State $B.set(C)$
\end{algorithmic}
\end{algorithm}

\cref{alg:datafecher} details the smart data prefetch. Let $R$ be the incoming request, $B_N$ the buffer name, $S_T$ the storage type, and content reference $C_R$ the input. Lines 1 and 2 create an empty buffer $B$ and storage client $SC$ respectively. Let $B_T$ be the buffer type in supported buffer collection $BC$. If buffer type $B_T$ is not empty and equals the buffer name $B_N$. Then assign buffer type $B_T$ to buffer instance $B$.
Let $S_E$ be storage in supported storage collection $SC$. If $S_E$ is not empty and $S_E$ equals the storage type $S_T$, then assign the specific storage client for storage type $S_T$ to $SC$.
In line 13, Truffle calls storage client $SC$ to get the data based on content reference $C_R$ input and assigns it to content $C$. Finally, set content $C$ in buffer $B$.

\subsection{Cold Start Pass Mechanism}\label{sec3:data_passing}

For data passing between two functions, Truffle first identifies where the source functions want to send the data. Then, once the target receives the data, Truffle stores it in a local buffer. 

\begin{figure}[!tb]
\centering
\hspace*{-0.3cm}
\includegraphics[width=.8\linewidth]{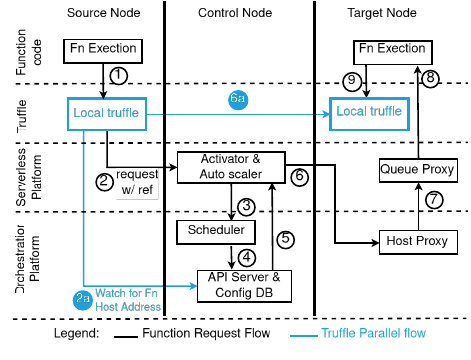}
\captionsetup{justification=centering}
\caption{Truffle Cold Start Pass}
\label{fig:truffle_datapass}
\end{figure}

\cref{fig:truffle_datapass} shows the communication between two serverless functions with its multiple platforms integration. \cref{fig:truffle_datapass} shows a simplified interaction between the orchestration and serverless platform for synchronous requests.
 
First, \textcircled{\small{1}} the source function fires a request to the Local truffle on the source. Then, \textcircled{\small{2}} Truffle accepts the request and identifies the target function based on information provided to the function. Then Truffle sends a request with an input data reference key to the serverless platforms such as Knative Activator. In \textcircled{\small{2a}}, simultaneously to \textcircled{\small{2}}, Truffle starts to listen for the target function host address. The function address is available after the scheduling phase. \textcircled{\small{3}} The platform activator receives the request and scales the function up via the orchestration tools mechanisms. 
\textcircled{\small{4}} Scheduling is finished and stored in the orchestration storage, e.g., Kubernetes etcd. \textcircled{\small{5}} Notifies the serverless platform. \textcircled{\small{6}} Resumes the request to the target Host Proxy. \textcircled{\small{6}} At this point, the Local Truffle on the source is aware of the target host reported by  \textcircled{\small{5}}, then simultaneously to \textcircled{\small{6}}, it starts transferring the outgoing input data to the Truffle buffer on the target host worker. In  \textcircled{\small{7}}, the Host Proxy forwards the request to the serverless platform ingress, such as Queue Proxy. In \textcircled{\small{8}}, the Queue Proxy forwards the request to the serverless function handler during Fn Execution. \textcircled{\small{9}} The function handler retrieves its input data from the Local truffle buffer with the input data reference key existent in the request from \textcircled{\small{2}}. 

\begin{algorithm}[!tb]
\caption{Truffle Cold Start Pass Request}
\label{alg:outgoing}
\begin{algorithmic}[1]
\Require $T_F$: target function 
\Require $T_C$: target content
\State $H_A \leftarrow \emptyset$
\State $W$ $\leftarrow$ resource watcher
\ForAll { $F (F \in F_E, \forall F_E \in W)$ }
    \If{$ F_H \ne \emptyset \land T_F = F_N$ }
        \State $H_A \leftarrow F_H$
    \EndIf
\EndFor
\State \textit{call\_target\_host\_truffle($H_A$,$T_C$)}
\end{algorithmic}
\end{algorithm}

Truffle Cold Start Data Pass relies on a data-passing request,~\cref{alg:outgoing} executed on the truffle instance on the source function host. Once the an incoming request, is executed in the target function host, the function retrieves the data from the buffer.
Let the target function $T_F$ and target content $T_C$ be the input. Let $F$ be function information in a function event $F_E$, where all function events belong to the watcher $W$. If function host information $F_H$ is not empty and target function $T_F$ equals the function event name $F_N$. Then, assign function host $F_H$ to target host address $H_A$. Once $H_A$ is known, stop listening for new events. Finally, call\_target\_host\_truffle with host address $H_A$ and target content $T_C$.

\section{Truffle Prototype Implementation} \label{sec4}

Truffle is published as an open-source framework part of the Polaris SLO CLoud. Polaris itself is part of the Linux Foundation Centaurus project. Truffle source code is available on GitHub\footnote{\url{https://github.com/polaris-slo-cloud/truffle}}. It is implemented in Go and currently supports Kubernetes as the orchestration tool and Knative as the Serverless Platform. 

Truffle runs as a DaemonSet on Kubernetes to ensure every node has an active instance and supports high availability with multiple instances per node, managed by a load balancer. Functions connect to Truffle through the HostIP and defined port. To improve latency, Truffle uses goroutines for parallel task execution, listening for events, and leveraging httputil Golang libraries to create a reverse proxy. Its data engine fetches data from remote services using external libraries, such as golang aws-sdk-go for AWS S3 and go-redis for KVS, and reads directly from the HTTP request body for direct communication. The Ingress component handles incoming requests, initiates parallel tasks with goroutines, creates a reference key, and forwards this to the Serverless platform, allowing the Serverless function to retrieve data from the local Truffle instance buffer. The Watcher monitors orchestration tools, specifically Kubernetes pod events via Kube API, enabling data transfer between functions immediately after a function is assigned to a node. 

\section{Evaluation}\label{sec5}
\subsection{Overview}
We design experiments to evaluate our proposed solutions with two Serverless workflows: Chained Functions and Video Analytics. Our Chained functions workflow is composed of two functions $a$ and $b$ that represents sequential execution of data-intensive functions. It receives certain input data and forwards it to the function. Our Video Analytics workflow is composed of the functions: \textit{Video Streaming}, \textit{Decoder}, and \textit{Image Recognition}  inspired by ~\cite{vHive,vSwarm} using most important invocation patterns from serverless computing fan-out and fan-in~\cite{CloudProgrammingSimplified}. We compare Truffle to three different data-passing storage types: Direct, Object Storage (S3), and Redis (KVS). As Truffle focuses on data-passing communication, we measure the time when the data is sent until the target functions receive the data.

To verify the proposed solution in this paper, we evaluate our framework by conducting experiments that measure data passing latency during the cold start. 

\paragraph{Metrics} Latency metric shows the time when a source function sends a request with a message until the target function receives it. This experiment includes also the entire target function startup including scheduling, cold start, and data transfer latency. Cold Start Delay metrics evaluate how cold start delays can influence the function's end-to-end execution time and how truffle can improve these delays to transfer large amounts of data. Therefore, we increase the function cold start in seconds and collect the function data passing latency (sec) for different input sizes.


\subsection{Experiment Setup}
We conduct experiments featuring the designed serverless workflow with AWS S3 and Redis KVS as the baseline. We use MicroK8s as the orchestration tool, and Knative as the Serverless Platform. Our microk8s cluster contains multiple nodes to ensure remote data passing experiments. For this purpose, we leverage Kubernetes node affinity to ensure the function scheduling on different nodes. Each cluster node is a VM with 4-core Intel Xeon Processor 2GHz, 8GB RAM with Ubuntu 22.04 LTS.
To avoid misleading results, we repeated the experiments and calculated the average result. Except for AWS S3, every tool was locally installed in the cluster with its vanilla version without configuration modifications including storage type Redis. For experiments with object object storage type, we leverage AWS S3. The same experiment setup is used to obtain the results for \cref{fig:coldstart} and \cref{fig:fh_lifecycle}

\subsection{Latency Results} \label{subsec:eval_latency}

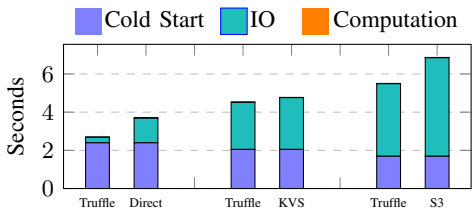
\begin{figure}[!b]
\centering
\begin{tikzpicture}[font=\small]
    \matrix[
        anchor=north ,
        nodes={inner sep=1pt} 
    ] at (0, 0) { 
        \node[fill=blue!50, minimum width=1em, minimum height=1em, align=center, baseline] {}; & 
        \node[baseline] {Cold Start}; & \hspace{0.5em} 
        \node[fill=cyan!70!yellow, draw=blue, minimum width=1em, minimum height=1em, align=center, baseline] {}; & 
        \hspace{1em} \node[baseline] {IO}; \hspace{2em}
        \node[fill=orange, minimum width=1em, minimum height=1em, align=center, baseline] {}; & 
        \hspace{6em} \node[baseline] {Computation}; \\
    };
\end{tikzpicture}

    \centering
    \begin{tikzpicture}[font=\small]
        \begin{axis}[
            ybar stacked,
            ymin=0,
            xtick=data,
            bar width=9pt,
            ylabel={Seconds},
            ymajorgrids=true,
            grid style=dashed,
            height=3.5cm,
            width=7cm,
            xticklabels={Truffle,Direct,,Truffle,KVS,,Truffle,S3},
            xticklabel style={font=\tiny},
            enlarge x limits=0.1, 
        ]
            \addplot [fill=blue!50]
                table [y=ColdStart, meta=Label, x expr=\coordindex]
                    {\testdata};
            \addplot [fill=cyan!70!yellow]
                table [y=IO, meta=Label, x expr=\coordindex]
                    {\testdata};
            \addplot [fill=orange]
                table [y=Computation, meta=Label, x expr=\coordindex]
                    {\testdata};
        \end{axis}
    \end{tikzpicture}
\caption{Total Workflow Execution Latency for 128MB, including function lifecycle phases}
\label{fig:latency-breakdown}
\end{figure}

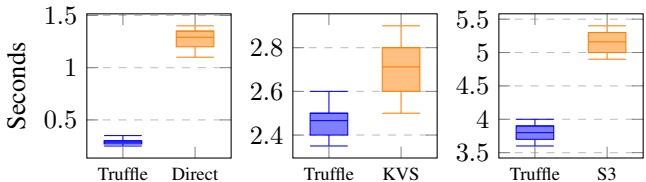
\begin{figure}[!b]
\centering
\begin{tikzpicture}
        \begin{axis} [
            box plot width=2.5mm,
            width=3.5cm,
            height=3.5cm,
            ylabel=Seconds,
            xtick={-1, 0},
            xticklabels={Truffle,Direct},
            ymajorgrids=true,
            grid style=dashed,
            xticklabel style={font=\scriptsize},
            enlarge x limits=0.5
        ]
        \boxplot [blue, fill=blue!50] {TruffleDirect.dat}
        \boxplot [color=orange, fill=orange!50] {Direct.dat}
        \end{axis}
\end{tikzpicture}
\begin{tikzpicture}
        \begin{axis} [
            box plot width=2.5mm,
            width=3.5cm,
            height=3.5cm,
            ylabel={},
            xtick={-1, 0},
            xticklabels={Truffle,KVS},
            ymajorgrids=true,
            grid style=dashed,
            xticklabel style={font=\scriptsize},
            enlarge x limits=0.5
        ]
        \boxplot [blue, fill=blue!50] {TruffleKVS.dat}
        \boxplot [color=orange, fill=orange!50] {KVS.dat}
        \end{axis}
\end{tikzpicture}
\begin{tikzpicture}
        \begin{axis} [
            box plot width=2.5mm,
            width=3.5cm,
            height=3.5cm,
            ylabel={},
            xtick={-1, 0},
            xticklabels={Truffle,S3},
            ymajorgrids=true,
            grid style=dashed,
            xticklabel style={font=\scriptsize},
            enlarge x limits=0.5
        ]
        \boxplot [blue, fill=blue!50] {TruffleS3.dat}
        \boxplot [color=orange, fill=orange!50] {S3.dat}
        \end{axis}
\end{tikzpicture}
\caption{IO latency impact on the total workflow execution for 128MB}
\label{fig:io_impact}
\end{figure}

In this experiment, we analyze how Truffle improves the total workflow latency when compared to the baselines. \cref{fig:latency-breakdown} presents the total workflow execution latency for Chained Functions while highlighting the function lifecycle phases. Due to Truffle's SDP that pre-fetches the data during the cold start, Truffle reduces the IO latency impact and consequently decreases the total workflow execution latency. More specifically, in \cref{fig:io_impact}, we show the IO latency impact on the total workflow latency. Truffle demonstrates an IO latency impact reduction of up to 77\%, improving the overall workflow execution latency by up to 46\% compared to the baselines: Direct, KVS and S3.

\pgfplotstableread{
Label ColdStart    IO Computation
Truffle  2.395     0.287    0.015
Direct   2.395     1.291   0.015

"" 0 0 0 

Truffle  2.049    2.466    0.012
KVS   2.049    2.712   0.012

"" 0 0 0 

Truffle  1.690    3.800    0.015
S3   1.690    5.160   0.015
}\testdata
    
\begin{figure*}[htb]
    \begin{subfigure}[b]{.23\textwidth}
        \begin{tikzpicture}[scale=.7]
          \begin{axis}[ 
            xlabel=Function input size in MB,
            ylabel={Latency (sec)},
            legend pos=north west,
            ymajorgrids=true,
            xmajorgrids=true,
            grid style=dashed,
            height=5cm,
            width=7cm
          ] 
          \addplot[smooth,mark=*,blue,ultra thick] coordinates
              {(1,4.208) (2,4.380) (4,4.435) (10,4.460) (20,4.697)(40,5.163)};
           \addlegendentry{Truffle}
            
           \addplot[smooth,mark=x,orange,ultra thick] coordinates
              {(1,4.363) (2,4.524) (4,4.509) (10,4.584) (20,4.935)(40,5.495)};
           \addlegendentry{Direct}

          \end{axis}
        \end{tikzpicture}
        \caption{Direct Data Passing}
        \label{fig:video_a}
    \end{subfigure}%
    \hspace*{2em}
    \begin{subfigure}[b]{.23\textwidth}
        \begin{tikzpicture}[scale=.7]
          \begin{axis}[ 
            xlabel= Function input size in MB,
            ylabel={},
            legend pos=north west,
            ymajorgrids=true,
            xmajorgrids=true,
            grid style=dashed,
            height=5cm,
            width=7cm
          ] 
          \addplot[smooth,mark=*,blue,ultra thick] coordinates
              {(1,3.954) (2,3.817) (4,4.101) (10,3.968) (20,5.223)(40,8.631)};
           \addlegendentry{Truffle+KVS}
           \addplot[smooth,mark=x,orange,ultra thick] coordinates  {(1,4.637) (2,4.375) (4,4.464) (10,5.222) (20,6.972) (40,10.315)};
           \addlegendentry{KVS}
          \end{axis}
        \end{tikzpicture}
        \caption{KVS Data Passing}
        \label{fig:video_b}
    \end{subfigure}%
    \hspace*{1em}
    \begin{subfigure}[b]{.23\textwidth}
    \begin{tikzpicture}[scale=.7]
      \begin{axis}[ 
        xlabel=Function input size in MB,
        ylabel={},
        legend pos=north west,
        ymajorgrids=true,
        xmajorgrids=true,
        grid style=dashed,
        height=5cm,
        width=7cm
      ] 
      \addplot[smooth,mark=*,blue,ultra thick] coordinates
          {(1,3.958) (2,3.742) (4,3.928) (10,3.733) (20,4.213) (40,5.712)};
       \addlegendentry{Truffle+S3}
      \addplot[smooth,mark=x,orange,ultra thick] coordinates
          {(1,5.568) (2,5.537) (4,5.769) (10,5.963) (20,6.175) (40,7.536)};
       \addlegendentry{S3}
      \end{axis}
    \end{tikzpicture}
    \caption{S3 Data Passing}
    \label{fig:video_c}
    \end{subfigure}%
\hspace*{-0.2em}
\begin{subfigure}[b]{.25\textwidth}
    \begin{tikzpicture}[scale=.7]
        \begin{axis}[
            ybar,   
            ymin=0,
            symbolic x coords={1,2,3,4,5},
            xtick=data,
            bar width=12pt,
            xlabel={\phantom{Label}},
            ylabel={Percentage(\%)},
            ylabel style={yshift=-10pt}, 
            ymajorgrids=true,
            grid style=dashed,
            height=5cm,
            width=6cm,        
            xticklabels={Direct Truffle,KVS Truffle,S3
            Truffle},
            xticklabel style={font=\scriptsize},
            enlarge x limits=0.2
        ]
        \addplot[
            fill=orange, 
            nodes near coords,
            every node near coord/.append style={
                font=\scriptsize, 
                white, 
                rotate=90,
                xshift=-50pt, 
                yshift=-6pt  
            }
        ] 
        coordinates {
            (1, 100) (3, 100) (5, 100)
        };
        

        \addplot[
            fill=blue, 
            nodes near coords,
            every node near coord/.append style={font=\scriptsize, white, rotate=90, xshift=-40pt, yshift=-6pt}
        ] 
        coordinates {
            (1, 93.9) (3, 83.6) (5, 75.8)
        }; 
        
        \end{axis}
    \end{tikzpicture}
    \caption{Normalized Latency at 40MB}  
    \label{fig:video_normalized}
\end{subfigure}    
\caption{Video Analytics Latency}
\label{fig:video_latency}
\end{figure*}
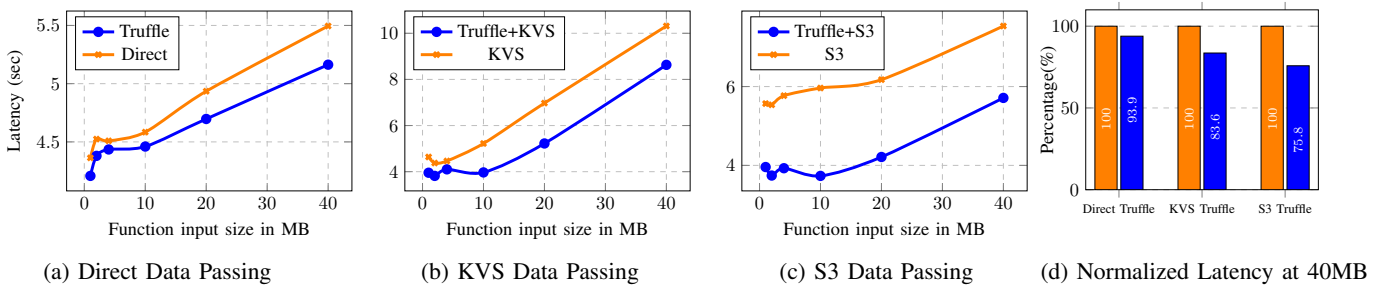

Further in \cref{fig:exp_latency} and \cref{fig:video_latency}, we use http synchronous serverless invocation trigger. We evaluate two applications: Chained Functions and Video Analytics. We increase the input size to validate the solution with different loads. \cref{fig:exp_latency} shows latency results for Chained Functions and \cref{fig:video_latency} shows latency results for Video Analytics workflow. \cref{fig:exp_latency} shows three different data passing strategies: direct, KVS, and object storage AWS S3. The function input size is represented on axis $x$ in MB and latency is shown on axis $y$ in seconds. 

\cref{fig:chained_a} shows that Truffle has latency from 1.852 sec to 2.697 sec while Direct HTTP requests show from 2.329 sec to 4.353 sec. \cref{fig:chained_nomalized} shows an improvement of up to 46\% on the direct data passing via Truffle. In \cref{fig:chained_b}, Truffle with KVS has latency from 2.060 sec to 4.827 sec while vanilla KVS shows from 2.121 sec to 5.073 sec. Truffle shows an improvement of up to 5\% on the KVS data passing compared to the baseline (\cref{fig:chained_nomalized}).
\cref{fig:chained_c} indicates that Truffle with S3 shows latency between 1.838 sec and 6.210 sec, whereas S3 alone ranges from 2.711 sec to 7.560 sec. This shows an improvement of up to 18\% in data passing via Truffle (\cref{fig:chained_nomalized}).

In the Video Analytics use case in \cref{fig:video_latency}, \cref{fig:video_a} reveals that Truffle has latency from 4.208 sec to 5.163 sec, while Direct HTTP requests range from 4.363 sec to 5.495 sec, an improvement of up to 6\% in direct data passing with Truffle (\cref{fig:video_normalized}). In \cref{fig:video_b}, Truffle with KVS has latency between 3.954 sec and 8.631 sec, compared to KVS alone, which ranges from 4.637 sec to 10.315 sec. This represents an improvement of up to 16\% in data passing with Truffle, shown in \cref{fig:video_normalized}. While in \cref{fig:video_c} indicates that Truffle with S3 shows latency from 3.958 sec to 5.712 sec, while S3 alone ranges from 5.568 sec to 7.536 sec. Truffle demonstrates an enhancement of up to 24\% (\cref{fig:video_normalized}).

\begin{figure*}[htb]
    \begin{subfigure}[b]{.23\textwidth}
        \label{fig:chained_a}
        \begin{tikzpicture}[scale=.7]
          \begin{axis}[ 
            xlabel=Function input size in MB,
            ylabel={Latency (sec)},
            legend pos=north west,
            ymajorgrids=true,
            xmajorgrids=true,
            grid style=dashed,
            ymax=5,
            height=5cm,
            width=7cm
          ] 
          \addplot[smooth,mark=*,blue,ultra thick] coordinates
              {(1,1.852) (2,1.893) (4,1.894) (8,2.029) (10,1.911) (16,2.111)(20,2.107) (32,2.108) (40,2.361) (64,2.272) (128,2.697)};
           \addlegendentry{Truffle}
            \addplot[smooth,mark=x,orange,ultra thick] coordinates {(1,2.329) (2,2.275) (4,2.456) (8,2.465) (16,2.592) (32,2.953) (64,3.359) (128,4.353)};
           \addlegendentry{Direct}
          \end{axis}
        \end{tikzpicture}
        \caption{Direct Data Passing}
        \label{fig:chained_a}
    \end{subfigure}%
    \hspace*{2em}
    \begin{subfigure}[b]{.23\textwidth}
        \begin{tikzpicture}[scale=.7]
          \begin{axis}[ 
            xlabel= Function input size in MB,
            ylabel={},
            legend pos=north west,
            ymajorgrids=true,
            xmajorgrids=true,
            grid style=dashed,
            height=5cm,
            width=7cm
          ] 
          \addplot[smooth,mark=*,blue,ultra thick] coordinates
              {(1,2.060) (2,1.944) (4,1.953) (8,2.031) (16,2.062) (32,2.043) (64,2.713) (128,4.827)};
           \addlegendentry{Truffle+KVS}
           \addplot[smooth,mark=x,orange,ultra thick] coordinates  {(1,2.121) (2,2.063) (4,2.080) (8,2.143) (16,2.335) (32,3.004) (64,3.413) (128,5.073)};
           \addlegendentry{KVS}
          \end{axis}
        \end{tikzpicture}
        \caption{KVS Data Passing}
        \label{fig:chained_b}
    \end{subfigure}%
    \hspace*{1em}
    \begin{subfigure}[b]{.23\textwidth}
    \begin{tikzpicture}[scale=.7]
      \begin{axis}[ 
        xlabel=Function input size in MB,
        ylabel={},
        legend pos=north west,
        ymajorgrids=true,
        xmajorgrids=true,
        grid style=dashed,
        height=5cm,
        width=7cm
      ] 
      \addplot[smooth,mark=*,blue,ultra thick] coordinates
          {(1,1.838) (2,1.705) (4,2.024) (8,2.061) (16,2.214) (32,2.298) (64,3.914) (128,6.210)};
       \addlegendentry{Truffle+S3}
      \addplot[smooth,mark=x,orange,ultra thick] coordinates
          {(1,2.711) (2,3.031) (4,3.199) (8,3.306) (16,3.284) (32,4.406) (64,5.185) (128,7.560)};
       \addlegendentry{S3}
      \end{axis}
    \end{tikzpicture}
    \caption{S3 Data Passing}
    \label{fig:chained_c}
    \end{subfigure}%
\hspace*{-0.2em}
\begin{subfigure}[b]{.25\textwidth}
    \begin{tikzpicture}[scale=.7]
        \begin{axis}[
            ybar,   
            ymin=0,
            symbolic x coords={1,2,3,4,5},
            xtick=data,
            bar width=12pt,
            xlabel={\phantom{Label}},
            ylabel={Percentage(\%)},
            ylabel style={yshift=-10pt}, 
            ymajorgrids=true,
            grid style=dashed,
            height=5cm,
            width=6cm,        
            xticklabels={Direct Truffle,KVS Truffle,S3
            Truffle},
            xticklabel style={font=\scriptsize},
            enlarge x limits=0.2
        ]
         \addplot[
            fill=orange, 
            nodes near coords,
            every node near coord/.append style={font=\scriptsize, 
            white, 
            rotate=90,
            xshift=-50pt, 
            yshift=-6pt  
        }] coordinates {(1, 100) (3, 100) (5, 100)};
        
        \addplot[
            fill=blue, 
            nodes near coords,
            every node near coord/.append style={font=\scriptsize, white, rotate=90, xshift=-30pt, yshift=-6pt}
        ] coordinates {(1, 53.5)  (3, 95.1)  (5, 82.1)};
        \end{axis}
    \end{tikzpicture}
    \caption{Normalized Latency at 128MB}    
    \label{fig:chained_nomalized}
\end{subfigure}  
\caption{Chained Functions Latency}
\label{fig:exp_latency}
\end{figure*}
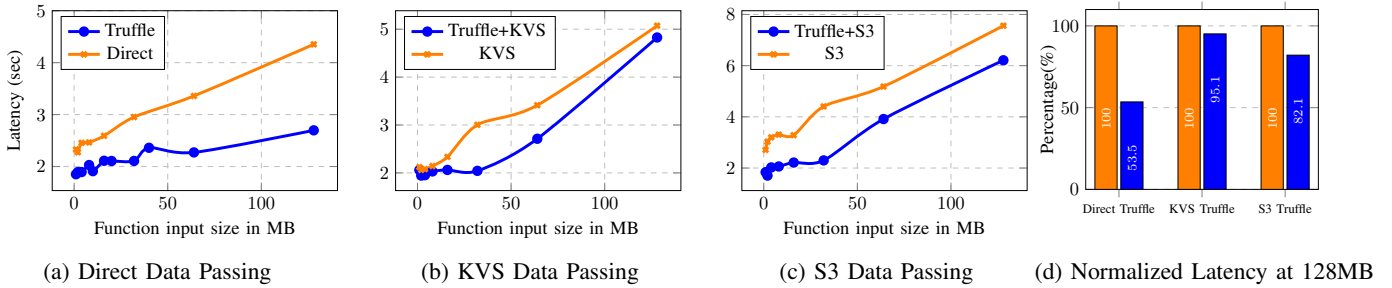

\subsection{Cold Start Delay Results} \label{subsec:eval_truffle_model}

\begin{figure}[htb]
\centering
\captionsetup[subfigure]{justification=centering}
\begin{subfigure}[t]{.23\linewidth}\centering
\begin{tikzpicture}[scale=.55,font=\large]
  \begin{axis}[ 
    xlabel=Added Delay (sec),
    ylabel={Latency (sec)},
    legend pos=north west,
    ymajorgrids=true,
    xmajorgrids=true,
    grid style=dashed,
    height=5cm,
    width=8cm
  ] 
  \addplot[blue,mark=*,ultra thick] coordinates
      {(0,5.934) (2,5.880) (4,6.041) (6,8.000) (8,9.810)(10,11.908)};
   \addlegendentry{Truffle+S3}
   \addplot[orange,mark=x,ultra thick] coordinates
      {(0,7.844) (2,8.562) (4,10.589) (6,12.563) (8,14.463)(10,17.175)};
   \addlegendentry{S3}
 
  \end{axis}
\end{tikzpicture}
\label{fig:coldstart_s3_a}
\end{subfigure}
\hspace*{\fill}%
\begin{subfigure}[t]{.23\linewidth}\centering
\begin{tikzpicture}[scale=.55,font=\large]
  \begin{axis}[
    xlabel=Added Delay (sec),
    ylabel={Latency (sec)},
    legend pos=north west,
    ymajorgrids=true,
    xmajorgrids=true,
    grid style=dashed,
    height=5cm,
    width=8cm
  ] 
  \addplot[blue,mark=*,ultra thick] coordinates
      {(0,1.654) (2,3.654) (4,5.686) (6,7.693) (8,9.833)(10,12.042)};
   \addlegendentry{Truffle+KVS}
   \addplot[orange,mark=x,ultra thick] coordinates
      {(0,3.763) (2,5.763) (4,7.811) (6,9.722) (8,11.792)(10,14.012)};
   \addlegendentry{KVS}
  \end{axis}
\end{tikzpicture}
\label{fig:coldstart_kvs_a}
\end{subfigure}
\hspace*{\fill}%
\caption{Additional Cold Start Delay with 100MB input}
\label{fig:coldstart_s3}
\end{figure}
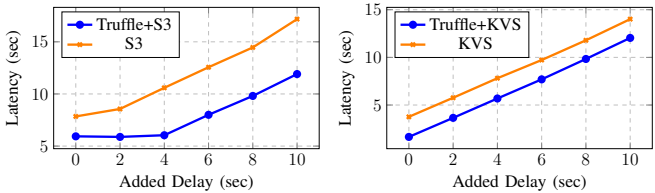

In this experiment, we compare the same storage approach in its conventional usage, i.e., after the function startup, and with Truffle, which leverages the cold start for fetching the input data. As Truffle leverages the cold start to fetch and pass data, the cold start latency directly impacts the function execution time and truffle performance. Thus, in this experiment, we increase the cold start latency for the different storage types, object storage AWS S3 and KVS. Truffle still leverages the same storage type as the given baseline. Furthermore, we add additional delay to the existing cold start to simulate longer cold start delays. Therefore, in  \cref{fig:coldstart_s3}, the baseline and truffle never intercept at $x = 0$ because $x$ represents the additional cold start. For example, $x=0$ in \cref{fig:coldstart_s3_a} is similar to ~\cref{fig:chained_c} $x=100 MB$. 
Hence, we evaluate how Truffle benefits applications with longer cold starts.

\cref{fig:coldstart_s3} shows the latency in seconds in axis $y$ and the cold start delay in axis $x$ using object object storage type AWS S3 and KVS. \cref{fig:coldstart_s3_a} shows the latency and cold start delay for an input size of 100MB while 
The orange line represents AWS S3 and KVS, and the blue line is Truffle + AWS S3. As applications only fetch the input data after the function has fully initialized, in \cref{fig:coldstart_s3_a}, the function latency (axis $y$) grows linear to the cold start delay $x$ in orange. On the other hand, in the blue line, Truffle delays the function latency increase. Truffle uses the full cold start to transfer the input data from 0 to 4s in axis $x$. Only at the second 6s in axis $x$ does the function latency start growing linearly.
Consequently, Truffle shows a latency decrease of up to 30\% in applications with longer cold starts. While applications with short cold start might profit 3\%.
In \cref{fig:coldstart_kvs_a}, KVS shows a latency from 3.7 seconds to 14 seconds in orange while Truffle shows from 1.6 seconds to 12 seconds in blue, which means Truffle decreases the latency up to 43\%. In \cref{fig:coldstart_kvs_a}, we observe that the input data transfer is faster than any cold start delay, which means even the shortest cold start delay is already sufficient to transfer the input data completely, which means Truffle optimizes execution time even when data transfer is fast enough to mask the cold starts. Therefore, this experiment shows similar linear growth regardless of the cold start delay.
\subsection{Discussion}

Truffle enhances serverless function data transfer and reduces latency up to 46\% by effectively utilizing cold start delays for data fetching. Therefore, it optimizes function initialization times, which is especially beneficial for data-intensive applications. Additionally, Truffle can integrate with different storage types and orchestrators via Truffle adapters. Furthermore, Truffle proposes function data fetching from different storage types, such as object storage, KVS, or direct passing. 
Truffle model \cref{subsec:truffle_model} shows that we can minimize function execution time by taking the shortest path which is either data transfer duration or cold start. Effectively, it enables us to overlap communication with cold start. Truffle utilizes this model and reduces the end-to-end latency by overlapping data transfer with the cold start. 

As discussed in \cref{subsec:truffle_model}, functions with longer cold starts profit more from Truffle than those with shorter cold starts. Results in \cref{subsec:eval_truffle_model} show that functions with longer cold starts profit up to 30\% more than functions with shorter cold starts. Truffle leverages the cold start for data fetching and transfer, which would otherwise be idle during the function lifecycle. Thus, small input data sizes do not significantly improve as functions with larger data input sizes.

Truffle's data transfer optimization reduces latencies and decreases function execution times for data-intensive tasks with a cold start. Therefore, during hot function executions, i.e., functions already running, Truffle checks whether the functions are running and forwards the data to the Serverless platform, acting as a transparent proxy. In such cases, Truffle's decision-making process during runtime is a trade-off. In the future, we plan to optimize Truffle's proxy for hot function to address this technical limitation.
In our evaluation, discussed in \cref{sec5}, we measure and analyze Truffle's improvement, i.e., \textit{Latency} for a serverless function with a cold start and \textit{Cold Start} duration, which shows how functions with long cold starts benefit from Truffle compared to those with short cold starts. We did not conduct throughput experiments to measure the data volume Truffle can handle, as it requires executing many requests in parallel. Therefore, throughput experiments only measure hot functions that do not contain any Truffle improvement.
Furthermore, we have conducted experiments with real-world applications to demonstrate broad applicability. However, the specific workflows we used in our experimental evaluation (Chained Functions and Video Analytics) may only represent some of the possible use cases for serverless computing. 
Moreover, our experiments use default configurations for the external systems such as Kubernetes and Knative which means any customized configuration that affects the performance of these systems might impact Truffle's result. Additionally, we compare Truffle with three different data-passing storage types: Direct, Object Storage (S3), and Redis (KVS), using the default configuration. The choice of these particular storage types and their default configurations may impact the results, as our experiments rely on the performance of these external services. Thus, any potential customized configuration affecting the system, network latency, or performance issues with S3 or Redis could affect our results. 

\section{Related Work} \label{sec6}
\textit{Cold Starts.} SEUSS~\cite{SEUSS} suggests creating and caching locally a snapshot to skip repetitive cold start steps, e.g., language runtime download and setup. Xanadu~\cite{Xanadu} reduces cascading cold start overheads by using speculative and just-in-time resource provisioning in a workflow. However, Xanadu focuses only the cold start mitigation which means functions are still responsible for fetching its input data. 
Nevertheless, SEUSS does not address the data passing which contributes to the function latency significantly. 
Faasm~\cite{faasm}, CWASI~\cite{Cwasi2023} and SAND~\cite{sand} follow a similar approach to sandbox sharing. These solutions propose to reuse a single sandbox for highly trusted functions, such as functions from the same workflow. 
As functions are executed in a single sandbox, there is only a simple cold start. Hence, reducing cold start from n functions in the workflow to a single cold start process. Nevertheless, sandbox sharing reduces the isolation and security provided by containerization. 

Truffle does not reduce the cold start latency but instead leverages it to optimize data transfer. It achieves this by integrating SDP and CSP, reducing overall latency, and improving serverless applications' efficiency. Cold start optimization approaches are not seen as contenders but rather as approaches that could be used in conjunction with Truffle.


\textit{Function Data Handling.}
 Lambdata~\cite{Lambdata} enables Serverless functions to create a cache of the function data so that subsequent requests and scheduling are prioritized based on the function and data locality. 
Although Lambdata has data locality in its scheduling, it does not consider the cold start. Thus, they do not implement input data fetching.
SONIC~\cite{sonic} relies on local storage and remote object storage to exchange data. For co-located functions, it exchanges data via local storage. For remote inter-function communication, Sonic leverages a metadata manager to send function references e.g. IP and Path. For persistent storage, Sonic also supports object storage such as AWS S3. 

Truffle improves the function of data handling by introducing the Smart Data Prefetch (SDP) mechanism, which automatically prefetches and transfers data. It reduces dependence on third-party services and abstracts data handling from the developer.

\textit{Inter-function Communication.}
 CloudBurst~\cite{cloudburstSF} relies on a local cache on each function host to allow low latency access to frequent data from a remote KVS. Although remote KVS is low latency and highly scalable KVS, it proposes data exchange via remote third-party service. 
Pocket~\cite{pocket} introduces elastic ephemeral storage for intermediate data exchange for serverless functions. Pocket in-memory storage offers scalability and improves communication compared to traditional object storage such as S3. Nevertheless, it still relies on an additional storage solution. 
XDT~\cite{xdt} creates direct communication between the functions by buffering the content on the sender and shipping a content key and the sender address. 
However, XDT does not consider the function cold start neither different storage types for the function data fetching, limiting its approach to inter-function communication. 

Truffle improves inter-function communication by implementing the Cold Start Pass (CSP) mechanism. This mechanism leverages the cold start to transfer data between functions, simplifying data transfer processes. Truffle integrates CSP with components such as Ingress and Buffer components, enhancing overall workflow efficiency and reducing latency in serverless applications.




\section{Conclusion \& Future Work} \label{sec7}

In this paper, we present Truffle, a novel architecture that enables serverless functions to execute parallel tasks to the cold start and consequently optimize the overall function end-to-end execution time.
Our experimental evidence indicates that the longest tasks during overall function execution are cold start and data transfer. Moreover, our analysis reveals that every task in the state-of-the-art serverless platform is executed sequentially, even when not dependent on one another. 

To address this issue, we have designed Truffle, which facilitates efficient data passing among Serverless functions. Truffle enables input data fetching and inter-function data passing during the cold start, i.e., ahead of the function execution. Furthermore, Truffle abstracts the function input data fetching for multiple storage types and optimizes the end-to-end function execution time. 

We evaluated Truffle by running different real-world Serverless applications, chained functions, and video analytics. Our experiments show that Truffle improves latency up to 46\% compared to the vanilla state-of-the-art Serverless platforms. The experiments also show that Serverless workflows with longer cold start delays profit more from Truffle, emphasizing its significance for applications that demand longer initialization, such as machine learning or not optimized language runtimes. Functions with longer cold starts of up to 10s profit nearly 30\% more than functions with shorter cold starts of up to 2s. 

In the current state-of-the-art sequential lifecycle phase execution, important tasks to the function are rarely applied due to increased function latency. We plan to leverage Truffle to execute tasks important to the Serverless function, such as serialization, deserialization, and argument validation. Serverless functions can potentially leverage Truffle Buffer to store its state handling and leverage the cold start to perform data-related tasks such as serialization without increasing the latency.
\section*{Acknowledgment}
This work is partially funded by the Austrian Research Promotion Agency (FFG) under the project RapidREC (Project No. 903884).

This research received funding from the EU’s Horizon Europe Research and Innovation Program under Grant Agreement No. 101070186. EU website for TEADAL: \url{https://teadal.eu}.

\printbibliography

\end{document}

%% file: plots/pgplots-definitions.tex
\pgfplotsset{
    box plot/.style={
        /pgfplots/.cd,
        black,
        only marks,
        mark=-,
        mark size=\pgfkeysvalueof{/pgfplots/box plot width},
        /pgfplots/error bars/y dir=plus,
        /pgfplots/error bars/y explicit,
        /pgfplots/table/x index=\pgfkeysvalueof{/pgfplots/box plot x index},
    },
    box plot box/.style={
        /pgfplots/error bars/draw error bar/.code 2 args={%
            \draw  ##1 -- ++(\pgfkeysvalueof{/pgfplots/box plot width},0pt) |- ##2 -- ++(-\pgfkeysvalueof{/pgfplots/box plot width},0pt) |- ##1 -- cycle;
        },
        /pgfplots/table/.cd,
        y index=\pgfkeysvalueof{/pgfplots/box plot box top index},
        y error expr={
            \thisrowno{\pgfkeysvalueof{/pgfplots/box plot box bottom index}}
            - \thisrowno{\pgfkeysvalueof{/pgfplots/box plot box top index}}
        },
        /pgfplots/box plot
    },
    box plot top whisker/.style={
        /pgfplots/error bars/draw error bar/.code 2 args={%
            \pgfkeysgetvalue{/pgfplots/error bars/error mark}%
            {\pgfplotserrorbarsmark}%
            \pgfkeysgetvalue{/pgfplots/error bars/error mark options}%
            {\pgfplotserrorbarsmarkopts}%
            \path ##1 -- ##2;
        },
        /pgfplots/table/.cd,
        y index=\pgfkeysvalueof{/pgfplots/box plot whisker top index},
        y error expr={
            \thisrowno{\pgfkeysvalueof{/pgfplots/box plot box top index}}
            - \thisrowno{\pgfkeysvalueof{/pgfplots/box plot whisker top index}}
        },
        /pgfplots/box plot
    },
    box plot bottom whisker/.style={
        /pgfplots/error bars/draw error bar/.code 2 args={%
            \pgfkeysgetvalue{/pgfplots/error bars/error mark}%
            {\pgfplotserrorbarsmark}%
            \pgfkeysgetvalue{/pgfplots/error bars/error mark options}%
            {\pgfplotserrorbarsmarkopts}%
            \path ##1 -- ##2;
        },
        /pgfplots/table/.cd,
        y index=\pgfkeysvalueof{/pgfplots/box plot whisker bottom index},
        y error expr={
            \thisrowno{\pgfkeysvalueof{/pgfplots/box plot box bottom index}}
            - \thisrowno{\pgfkeysvalueof{/pgfplots/box plot whisker bottom index}}
        },
        /pgfplots/box plot
    },
    box plot median/.style={
        /pgfplots/box plot,
        /pgfplots/table/y index=\pgfkeysvalueof{/pgfplots/box plot median index}
    },
    box plot width/.initial=1em,
    box plot x index/.initial=0,
    box plot median index/.initial=1,
    box plot box top index/.initial=2,
    box plot box bottom index/.initial=3,
    box plot whisker top index/.initial=4,
    box plot whisker bottom index/.initial=5,
}

\newcommand{\boxplot}[2][]{
    \addplot [box plot median,#1] table {#2};
    \addplot [forget plot, box plot box,#1] table {#2};
    \addplot [forget plot, box plot top whisker,#1] table {#2};
    \addplot [forget plot, box plot bottom whisker,#1] table {#2};
}